\documentstyle[12pt]{article}

\newcommand{\greaterthansquiggle}{\raise.3ex\hbox%
                        {$>$\kern-.75em\lower1ex\hbox{$\sim$}}}
\newcommand{\lessthansquiggle}{\raise.3ex\hbox%
                        {$<$\kern-.75em\lower1ex\hbox{$\sim$}}}
\newcommand{\be}{\begin{equation}}
\newcommand{\ee}{\end{equation}}
\newcommand{\ba}{\begin{eqnarray}}
\newcommand{\ea}{\end{eqnarray}}
\newcommand{\no}{\nonumber}

\newcommand{\lets}{\lessthansquiggle}
\newcommand{\mixing}{$D^0$--$\bar D^0$ mixing}
\newcommand{\system}{$D^0$--$\bar D^0$ system}
\newcommand{\ddbar}{$D^0/\bar D^0$ }
\newcommand{\ann}{a(f)}
\newcommand{\abn}{\bar a(f)}
\newcommand{\anb}{a(\bar f)}
\newcommand{\abb}{\bar a(\bar f)}
\newcommand{\bnn}{b(f)}
\newcommand{\bbn}{\bar b(f)}
\newcommand{\bnb}{b(\bar f)}
\newcommand{\bbb}{\bar b(\bar f)}
\newcommand{\cnn}{c(f)}
\newcommand{\cbn}{\bar c(f)}
\newcommand{\cnb}{c(\bar f)}
\newcommand{\cbb}{\bar c(\bar f)}
\normalsize
\begin{document}

\bibliographystyle{plain}

\begin{titlepage}

\begin{flushright}
UWThPh-1996-26 \\
FISIST/3-96/CFIF \\
\today \\
\end{flushright}

\vspace{2cm}

\begin{center}

{\Large \bf On the Determination of $\Delta m$ and $\Delta \Gamma$\\[10pt]
in Tagged \ddbar  Decays}\\[70pt]

W. Grimus\\
Institut f\"ur Theoretische Physik, Universit\"at Wien\\
Boltzmanngasse 5, 
A-1090 Wien, Austria\\[6mm]
L. Lavoura \\
Universidade T\'ecnica de Lisboa \\
CFIF, Instituto Superior T\'{e}cnico,
Edif\'{\i}cio Ci\^{e}ncia (f\'{\i}sica) \\
P-1096 Lisboa Codex, Portugal

\vspace{2.5cm}

{\bf Abstract}\\[7pt]

\end{center}

We consider the time dependence of the decays
of tagged $D^0$ and $\bar{D}^0$
into CP-conjugate final states $f$ and $\bar{f}$,
or into a CP eigenstate $F$.
We expand each decay width as $\exp (- \Gamma t)$ times
a series in $\Gamma t$, where $\Gamma$ is the average decay width of the
mass eigenstates $D_H$ and $D_L$,
and examine the first three terms of the series.
We show that experimental information on the coefficients of 
these terms allows in principle to compute all the relevant
mixing parameters.
In particular, depending on CP violation or conservation,
we discuss the different possibilities 
to extract $\Delta m$ and $\Delta \Gamma$, i.e.,
the mass and decay-width differences of the mass eigenstates.
We also comment on
consistency conditions among the coefficients
of the various decay-width expansions.

\end{titlepage}

In this paper we consider the system of the charmed neutral mesons
$D^0$ and $\bar D^0$.
The mass eigenstates of that system are given by
\ba \label{kets}
| D_H \rangle & = & p |D^0 \rangle + q |\bar D^0 \rangle\, ,
\no \\
| D_L \rangle & = & p |D^0 \rangle - q |\bar D^0 \rangle\, ,
\ea
where the index $H$ refers to heavy and $L$ to light.
We denote by
$ \Gamma = (\Gamma_H + \Gamma_L) / 2 $
the average decay width of $ D_H $ and $ D_L $.
The standard model (SM) predicts very small parameters
\be \label{definexy}
x=\frac{m_H-m_L}{\Gamma}
\quad \mbox{and} \quad
y=\frac{\Gamma_H-\Gamma_L}{2\Gamma}\, .
\ee
The parameter $x$ is positive by definition.
Experimentally,
it is already known that $ (x^2+y^2)\, \lets\, 10^{-2}$ \cite{part}.
Therefore,
even if physics beyond the SM is very important in the \system,
at most the onset of oscillations can be discovered \cite{big}.
This fact is exploited in discussions of tagged \ddbar  decays
with decay-time information,
where in the time-dependent decay widths
one performs an expansion \cite{big,bla,wol}
with respect to the quantity $(x-iy) \Gamma t$,
which is small as long as $\Gamma t$ is of order one. 
It is reasonable to truncate the expansion at order $(\Gamma t)^2$.
There is some hope that in future experiments the coefficients
of this expansion will be measured.
In this paper we propose to use these coefficients
to get information on $x$ and $y$ and also on CP violation.
We show that if one measures them up to order $(\Gamma t)^2$
in all four decay widths
$\Gamma (D^0(t)/\bar D^0(t) \rightarrow f/\bar f)$
of a given Cabibbo-allowed/doubly-Cabibbo-suppressed
pair $f/\bar f$ of CP-conjugate final states
one can unambiguously extract $x$,
$y$,
and also information on CP violation in mixing.
The simplest such final state would be $f=K^-\pi^+$.
We also include a discussion of final CP eigenstates $F$
like $F=K^+K^-$.

To study phenomenologically the time dependence of the widths
of tagged $D^0$ and $\bar D^0$ decays
into $f$ and $\bar f$
the following four amplitudes are relevant:
\be \label{amplitudes}
{\cal A}(D^0 \rightarrow f) \equiv A, \
{\cal A}(\bar D^0 \rightarrow \bar{f}) \equiv \bar A, \
{\cal A}(\bar D^0 \rightarrow f) \equiv B, \
{\cal A}(D^0 \rightarrow \bar{f}) \equiv \bar B.
\ee
As mentioned above the starting point of our analysis
is an expansion of the
time-dependent \ddbar decay widths with
respect to $(x-iy) \Gamma t$.
The first three terms in this expansion are given by 
\be \label{expansion}
\Gamma(\stackrel{(-)}{D^0}(t) \rightarrow \stackrel{(-)}{f}) =
\mbox{e}^{-\Gamma t} [ \stackrel{(-)}{a}(\stackrel{(-)}{f})\, +
\stackrel{(-)}{b}(\stackrel{(-)}{f}) \,\Gamma t\, +
\stackrel{(-)}{c}(\stackrel{(-)}{f}) \,(\Gamma t)^2 + \ldots
]
\ee
with the coefficients
$$
\begin{array}{cc}
\ann = |A|^2, \;& \bnn = \mbox{Im} [ (x-iy) \frac{q}{p} A^* B ], \\[2pt]
\abn = |B|^2, \;& \bbn = \mbox{Im} [ (x-iy) \frac{p}{q} B^* A ], \\[2pt]
\anb = |\bar B|^2, \;& \bnb = \mbox{Im} [ (x-iy) \frac{q}{p} 
\bar B^* \bar A ], \\[2pt]
\abb = |\bar A|^2, \;& \bbb = \mbox{Im} [ (x-iy) \frac{p}{q} 
\bar A^* \bar B ], 
\end{array}
$$
\be \label{coeff} \begin{array}{c}
\cnn = \frac{1}{4} [ \eta^2 |B|^2 (x^2+y^2)-|A|^2 (x^2-y^2) ], \\[2pt]
\cbn = \frac{1}{4} [ \frac{1}{\eta^2} |A|^2 (x^2+y^2)-|B|^2
(x^2-y^2) ], \\[2pt]
\cnb = \frac{1}{4} [ \eta^2 |\bar A|^2 (x^2+y^2)-|\bar B|^2
(x^2-y^2) ], \\[2pt]
\cbb = \frac{1}{4} [ \frac{1}{\eta^2} |\bar B|^2 (x^2+y^2)-
                                      |\bar A|^2 (x^2-y^2) ],
\end{array}
\ee
where $\eta \equiv |q/p|$ parametrizes T and CP violation in \mixing.
We assume that $\Gamma$ and the coefficients in eqs.~(\ref{coeff})
can be obtained from a fit of eq.~(\ref{expansion}) to the \ddbar decays
with decay-time information.

Let us first assume that we have the coefficients 
$\stackrel{(-)}{a}(\stackrel{(-)}{f})$ and 
$\stackrel{(-)}{c}(\stackrel{(-)}{f})$ at our disposal.
This allows us to compute $x$ and $y$ as functions of these coefficients
and of $\eta$:
\ba \label{xy}
x^2 & = & 2 \,\frac{\eta^2 \cbn - \cnn}{\ann - \eta^2 \abn} = 
       2 \,\frac{\eta^2 \cbb - \cnb}{\anb - \eta^2 \abb}\, ,
               \no \\
y^2 & = & 2 \,\frac{\eta^2 \cbn + \cnn}{\ann + \eta^2 \abn} = 
       2 \,\frac{\eta^2 \cbb + \cnb}{\anb + \eta^2 \abb}\, .
\ea
This means that given $\eta$, then $x$ and $y$ can be obtained from
considering tagged \ddbar  decays into either $f$ or $\bar f$.
If the full information is used the consistency condition
\be \label{condition}
\cnn \abb + \cbn \anb = \cbb \ann + \cnb \abn
\ee
is obtained.
Similarly,
$\eta$ can be extracted by
\be \label{eta}
\eta^4 = \frac{\cnb \ann - \cnn \anb}{\cbn \abb - \cbb \abn}\, .
\ee
It is important to note that the determination of $\eta$
depends not only on the observation
of the shape of the decay curves of $D^0(t)$ and $\bar D^0(t)$,
but also on the {\em relative normalization} of these decay curves.
This means that it is not just the ratios $ \cnb : \cnn : \anb : \ann $
and $ \cbn : \cbb : \abn : \abb $ which are important;
for the computation of $\eta$ the relative normalization
$ \ann : \abn $ is fundamental too.
This is a point relevant for the observation of T violation
in the mixing of any neutral-meson system \cite{luis}. It is
easy to check by inserting eq.~(\ref{eta}) into eqs.~(\ref{xy}) that
the knowledge of this relative normalization is not necessary for the
determination of $x$ and $y$.

Let us now assume instead that the coefficients
$\stackrel{(-)}{a}(\stackrel{(-)}{f})$ and
$\stackrel{(-)}{b}(\stackrel{(-)}{f})$ are available.
Then,
we can make a simple parameter counting.
If we add $\eta$ to those eight coefficients,
there is a total of nine measurements.
On the other hand,
only seven rephasing--invariant quantities
can be formed from $q/p$ and from the amplitudes
in eq.~(\ref{amplitudes}) \cite{BGL}.
Therefore,
there are two consistency conditions involving the parameters $x$ and $y$,
which are thus calculable.
These conditions are
\be \label{e1}
\frac{1}{x^2} \left[ \bnn - \eta^2 \bbn \right]^2 +
\frac{1}{y^2} \left[ \bnn + \eta^2 \bbn \right]^2 = 4 \ann \abn \eta^2
\ee
and
\be \label{e2}
\frac{1}{x^2} \left[ \bnb - \eta^2 \bbb \right]^2 +
\frac{1}{y^2} \left[ \bnb + \eta^2 \bbb \right]^2 = 4 \abb \anb \eta^2,
\ee
which are equations for ellipses in the variables $1/x$ and $1/y$.
Eq.~(\ref{e2}) is obtained from eq.~(\ref{e1})
by simply replacing $f$ by $\bar f$.
>From the system of eqs.~(9) and (10) one can find $x$
(which is positive by definition)
and $|y|$.
This method for the determination of the mass and decay-width differences
only works,
however,
if $\eta$ is already known.
This could be from eq.~(\ref{eta}),
for instance.
If we combine eqs.~(\ref{e1}) and (\ref{e2}) and
eqs.~(\ref{xy}) we get
two further consistency conditions analogous to eq.~(\ref{condition}).

If CP is conserved we have $\eta = 1$ and the relations
\be \label{CPcondition}
\ann = \abb, \: \abn = \anb, \: \bnn = \bbb, \:
\bnb = \bbn, \: \cnn = \cbb, \: \cnb = \cbn
\ee
are fulfilled.
Then the two ellipses coincide and $x$ and $y$ cannot be disentangled.
Therefore, eqs.~(\ref{e1}) and (\ref{e2}) constitute another version
of the observation that large CP violation can
be quite helpful in getting a hold on mixing in the \system\ \cite{wol}.
Nevertheless, even in the case of CP conservation
one could have an interesting restriction
in the $x-y$ plane,
and consequently lower bounds on $x$ and $y$.

The previous method using $\stackrel{(-)}{c}(\stackrel{(-)}{f})$ 
does not suffer from the above problem,
i.e.,
$x$ and $y$ can be determined even when CP is conserved.
However,
in that case the two ways (see eqs.~(\ref{xy})) of
obtaining $x$ and $y$ via the final states $f$ and $\bar f$ coincide.

If CP is conserved the constraint eq.~(\ref{condition})
is automatically satisfied because of the equalities
in eqs.~(\ref{CPcondition}).
However,
CP conservation introduces an additional constraint
among the coefficients $\stackrel{(-)}{a}(f)$,
$\stackrel{(-)}{b}(f)$ and $\stackrel{(-)}{c}(f)$, (or,
analogously,
among the coefficients $\stackrel{(-)}{a}(\bar f)$,
$\stackrel{(-)}{b}(\bar f)$ and $\stackrel{(-)}{c}(\bar f)$)
which reads 
\be
4 a \bar a (\bar c^2 - c^2) =
(b^2 + \bar b^2) (a \bar c - \bar a c)
+ 2 b \bar b (\bar a \bar c - a c).
\ee
We have left out the labels $f$ or $\bar f$ for simplicity of notation.

We now proceed to an analogous discussion of \ddbar decays into a CP 
eigenstate $F$.
In this case the expansion of the decay widths reads
\be \label{expansionF}
\Gamma(\stackrel{(-)}{D^0}(t) \rightarrow F) =
\mbox{e}^{-\Gamma t} [ \stackrel{(-)}{a}(F)\, +
\stackrel{(-)}{b}(F) \,\Gamma t\, +
\stackrel{(-)}{c}(F) \,(\Gamma t)^2 + \ldots ]
\ee
with
$$
\begin{array}{cc}
a(F)=|A|^2, \;&
b(F)= \mbox{Im} [(x-iy) \frac{q}{p} A^* \bar A], \\[2pt]
\bar a(F)=|\bar A|^2, \;&
\bar b(F)= \mbox{Im} [(x-iy) \frac{p}{q} A \bar A^*],
\end{array}
$$
\be\begin{array}{c}
c(F) =
\frac{1}{4} [\eta^2 |\bar A|^2 (x^2+y^2) - |A|^2 (x^2-y^2)], \\[2pt]
\bar c(F) =
\frac{1}{4} [\frac{1}{\eta^2} |A|^2 (x^2+y^2) - |\bar A|^2 (x^2-y^2)].
\end{array}
\ee
In the following we shall leave out the label $F$ in the coefficients.
The expansion of eq.~(\ref{expansionF})
supplies us with six experimental quantities. 
This number has to be confronted with the number of independent
physical quantities constructed from $A=\bar B$,
$\bar A=B$ and $q/p$,
which is now four \cite{BGL}.
Therefore,
at least in principle,
one can obtain $x$ and $y$ from such measurements.
Analogous to eqs. (\ref{xy}) we find
\be \label{xyF}
x^2 = 2 \,\frac{\eta^2 \bar c - c}{a - \eta^2 \bar a}\, , \quad
y^2 = 2 \,\frac{\eta^2 \bar c + c}{a + \eta^2 \bar a}\, .
\ee
Of course,
here there is no consistency condition such as eq.~(7).
Also,
$\eta$ cannot be obtained from the coefficients
$\stackrel{(-)}{a}$ and $\stackrel{(-)}{c}$ alone.
On the other hand,
from $b$ and $\bar b$ we get a single ellipse
\be \label{ellipseF}
\frac{1}{x^2} (b -\eta^2 \bar b)^2 + \frac{1}{y^2} (b + \eta^2
\bar b)^2 = 4 a \bar a \eta^2
\ee
in the variables $1/x$ and $1/y$.
Taking eqs.~(\ref{xyF}) and (\ref{ellipseF}) together we can compute
\be \label{etaF}
\eta^4 = \frac{b^2 (\bar a c-a \bar c)+2b \bar b ac-4a \bar a
c^2}{\bar b^2(a \bar c-\bar a c)+2b \bar b \bar a \bar c-4a \bar
a \bar c^2}\, ,
\ee
and then from eqs.~(\ref{xyF}) the values of $x$ and $y$ can be obtained.

CP conservation is a much stronger restriction on the
phenomenology of decays into CP eigenstates $F$
than of decays into $f$ and $\bar f$,
because it gives not only
\be
a=\bar a, \; b=\bar b, \; c=\bar c,
\ee
but also
\be \label{y}
|b|=|y|a \quad \mbox{and} \quad c= \frac{1}{2} y^2 a,
\ee
and therefore
\be
2 a c = b^2.
\ee
In this case the expressions for $x$ and $\eta$
are of the type 0/0 and thus undefined.
As a consequence,
if CP violation in mixing and amplitudes is small,
it will be difficult to constrain $x$ and $\eta$ from the decay
of tagged \ddbar into CP eigenstates. 
However,
eq.~(\ref{y}) shows that CP eigenstates are well suited to get
information on $y$ \cite{liu}.
Using CP non-eigenstates one does not face the problem
of undefined expressions in the case of CP conservation.
The denominators in eqs.~(\ref{xy}) and (\ref{eta})
are different from zero,
because the amplitude $B$ is doubly-Cabibbo-suppressed,
i.e.,
$B \sim \lambda^2 A$ with $\lambda \approx 0.22$.

In conclusion,
we have studied the conditions under which
we can extract the mixing parameters $x$ and $y$ of the \system\
from the measurement of the first three terms in an expansion
in $(x-iy) \Gamma t$ of the decay widths
of tagged $D^0(t)$ and $\bar D^0(t)$.
We found that this task can indeed be performed in several ways
if one considers the decays into a pair of CP-conjugate final states.
If one uses the coefficients of $(\Gamma t)^0$ and $(\Gamma t)^2$
it suffices to consider either $f$ or $\bar f$ as a final state,
provided the parameter $\eta$ measuring CP violation in \mixing\
is determined from elsewhere.
If the coefficients of all four decay widths are given
then $\eta$ can be determined as well and,
in addition,
a consistency condition among the coefficients has to be fulfilled.
Experimental knowledge of the coefficients
of $(\Gamma t)^0$ and $(\Gamma t)^1$ for both final states $f$ and $\bar f$
allows the extraction of $y$,
but in order to obtain $x$, CP must be violated.
We have also discussed the decays into a CP eigenstate.
Again CP violation is necessary to determine the mixing parameters.
In any case,
in the framework discussed here,
the determination of $\eta$ seems to be
the most difficult experimental problem.
Needless to say,
all the different methods we put forward
to get information on \mixing\ can be combined in various ways.
The experimental situation will pin down the most useful strategy.

\vfill
\end{document}